\documentclass[aps,superscriptaddress,showpacs]{revtex4-2}

\usepackage{graphicx}
\usepackage{epstopdf}
\usepackage{caption}
\usepackage{color}
\usepackage{bm}
\usepackage{textcomp} 
\usepackage{subfigure}
\usepackage{ulem}
\usepackage{color}
\usepackage{mathtools}
\usepackage{amsmath}
\graphicspath{{figures/}}

\begin{document}

\title{Considerations on measuring spin-spin correlation of hyeprons in heavy-ion experiments}

\author{Diyu Shen}
\email{dyshen@fudan.edu.cn}
\affiliation{Key Laboratory of Nuclear Physics and Ion-beam Application (MOE), Institute of Modern Physics, Fudan University, Shanghai 200433, China}
\author{Jinhui Chen}
\affiliation{Key Laboratory of Nuclear Physics and Ion-beam Application (MOE), Institute of Modern Physics, Fudan University, Shanghai 200433, China}
\author{Aihong Tang}
\affiliation{Brookhaven National Laboratory, Upton, New York 11973}
\date{July 2024}

\begin{abstract}
     The significant global spin alignment observed for $\phi$ mesons in heavy-ion collisions has sparked intense discussions about its origin and implications. One explanation suggests that fluctuations in the strong force field may introduce strong spin correlations between strange ($s$) and anti-strange ($\bar{s}$) quarks, leading to the global spin alignment of $\phi$ mesons. Extending this line of research, the theoretical community has proposed studying the spin correlation between $\Lambda$ and $\bar{\Lambda}$ hyperons. In this paper,  we construct experimental observables and make connections between them and the theoretical proposed quantities.
    
\end{abstract}
\maketitle
 
 The concept of global polarization in heavy-ion collisions was first proposed in Ref.\cite{Liang:2004ph}. In off-central heavy-ion collisions, quarks are predicted to be globally polarized along the direction of angular momentum due to spin-orbital coupling in strong interactions. This effect has recently been confirmed experimentally through measurements of $\Lambda$ global polarization~\cite{STAR:2017ckg,STAR:2018gyt,ALICE:2019onw,STAR:2021beb,STAR:2023nvo}. However, another related phenomenon, vector meson spin alignment, cannot be explained solely by vorticity or other  mechanisms, most of which are conventional~\cite{Liang:2004xn,Becattini:2013vja, Yang:2017sdk,Sheng:2019kmk,Xia:2020tyd,Gao:2021rom,Muller:2021hpe}. The measured $\rho_{00}$, the 00-component of the spin density matrix (with 1/3 indicating no spin alignment), of $\phi$ mesons is significantly above 1/3~\cite{STAR:2022fan}, while predictions from the quark coalescence model, considering $\Lambda$ data, suggest~\cite{Liang:2004xn, Yang:2017sdk}  $\rho_{00}$ below 1/3 with a much smaller deviation from 1/3. This discrepancy between hyperon polarization and vector meson spin alignment implies richer spin physics beyond global quark polarization. It was later realized that spin correlation between quarks and anti-quarks plays a crucial role in determining vector meson spin~\cite{Lv:2024uev}. For instance, short-range correlations between $s$ and $\bar{s}$ quarks due to strong-force fields successfully explain the observed $\phi$ meson spin alignment~\cite{Sheng:2022wsy,Sheng:2019kmk,Sheng:2020ghv,Sheng:2022ffb,Kumar:2023ghs}.

The spin-spin correlation of spin 1/2 particles was usually studied in helicity-frame in $e^+e^-$ collisions or $p+p$ collisons~\cite{Chen:2016iey,Li:2023qgj}, and the correlation function is defined as 
\begin{equation}
c_{ij} = \frac{f_{\uparrow\uparrow} + f_{\downarrow\downarrow} - f_{\downarrow\uparrow} - f_{\uparrow\downarrow}}{f_{\uparrow\uparrow} + f_{\downarrow\downarrow} + f_{\downarrow\uparrow} + f_{\uparrow\downarrow}},
\label{Eq:cnn}
\end{equation}
where $f_{m_im_j}=\langle m_im_j | \hat{\rho} | m_im_j \rangle$ denotes the fraction of fermion pairs in the spin state of $\left | m_im_j  \right \rangle$ along spin quantization direction, with $\hat{\rho}$ being the spin density matrix. In general, the quantization directions of particle $i$ and $j$ differ in helicity-frame. However, in heavy-ion collisions, we are interested in global spin-spin correlations, meaning the quantization directions of particle $i$ and $j$ are fixed. In this case, Eq.~\ref{Eq:cnn} is modified as~\cite{Lv:2024uev}:
\begin{equation}
c'_{ij} = c_{ij} - P_i P_j,
\label{Eq:cnn_prime}
\end{equation}
where $P_i$ and $P_j$ are the polarization of particle $i$ and $j$ respectively. It is clear that the $c'_{ij}$ represents $\langle \hat{\sigma}_{m_i} \hat{\sigma}_{m_j} \rangle - \langle \hat{\sigma}_{m_i} \rangle \langle \hat{\sigma}_{m_j} \rangle$, where $\hat{\sigma}_m$ is the Pauli operator along spin quantization direction. 

The left-hand side of Eq.~\ref{Eq:cnn_prime} represents the global spin correlations in heavy-ion collisions and can be calculated by measuring the $c_{ij}$ and global polarization. In the following, we demonstrate how to measure $c_{ij}$ using the example of $\Lambda$-$\Lambda$ spin correlation. This method can be easily extended to other hyperons. 

We begin with the proton distribution from $\Lambda$ weak decay,
\begin{equation}
    \frac{dN}{d\cos\theta^*}=\frac{1}{2}(1+\alpha_\Lambda P_\Lambda\cos\theta^*),
\end{equation}
where $\theta^*$ is the angle, in $\Lambda$'s rest frame, between proton momentum and $\Lambda$ spin quantization direction. $P_\Lambda$ defines $\Lambda$ polarization and $\alpha_\Lambda$ is a decay constant. For $\Lambda$ pairs in the spin state $\left | \uparrow\uparrow  \right \rangle$ (both are fully polarized thus $P_\Lambda=1$), the distribution of proton-pairs can be written as
\begin{equation}
\frac{dN_{\uparrow\uparrow}}{d\cos\theta_i^*d\cos\theta_j^*}=\frac{1}{4}\left [1+\alpha_\Lambda (\cos\theta_i^*+\cos\theta_j^*) + \alpha_\Lambda^2\cos\theta_i^*\cos\theta_j^* \right ],
\label{Eq:Npp}
\end{equation}
where $N_{\uparrow\uparrow}$ denotes proton pairs from $\Lambda$ pairs in $\left | \uparrow\uparrow  \right \rangle$ state. The footnote $i$ and $j$ denote the first proton and the second proton, respectively. Similarly, for $\Lambda$ pairs in the spin state of $\left | \downarrow\downarrow  \right \rangle$, $\left | \uparrow\downarrow  \right \rangle$, and $\left | \downarrow\uparrow  \right \rangle$, the proton pairs are described by 
\begin{eqnarray}
\frac{dN_{\downarrow\downarrow}}{d\cos\theta_i^*d\cos\theta_j^*}&=&\frac{1}{4}\left [1-\alpha_\Lambda (\cos\theta_i^*+\cos\theta_j^*) + \alpha_\Lambda^2\cos\theta_i^*\cos\theta_j^* \right ], \label{Eq:Nnn}\\
\frac{dN_{\uparrow\downarrow}}{d\cos\theta_i^*d\cos\theta_j^*}&=&\frac{1}{4}\left [1+\alpha_\Lambda (\cos\theta_i^*-\cos\theta_j^*) - \alpha_\Lambda^2\cos\theta_i^*\cos\theta_j^* \right ], \label{Eq:Npn}\\
\frac{dN_{\downarrow\uparrow}}{d\cos\theta_i^*d\cos\theta_j^*}&=&\frac{1}{4}\left [1-\alpha_\Lambda (\cos\theta_i^*-\cos\theta_j^*) - \alpha_\Lambda^2\cos\theta_i^*\cos\theta_j^* \right ]. \label{Eq:Nnp}
\end{eqnarray}
In reality, $\Lambda$ pairs are mixture of the above spin states, with the fractions given by $f_{m_im_j}$. Therefore, the general distribution of daughter proton pairs can be expressed as follows :
\begin{eqnarray}
\frac{dN}{d\cos\theta_i^*d\cos\theta_j^*} &=& f_{\uparrow\uparrow}\frac{dN_{\uparrow\uparrow}}{d\cos\theta_i^*d\cos\theta_j^*} + f_{\downarrow\downarrow}\frac{dN_{\downarrow\downarrow}}{d\cos\theta_i^*d\cos\theta_j^*} + f_{\uparrow\downarrow}\frac{dN_{\uparrow\downarrow}}{d\cos\theta_i^*d\cos\theta_j^*} + f_{\downarrow\uparrow}\frac{dN_{\downarrow\uparrow}}{d\cos\theta_i^*d\cos\theta_j^*} \notag \\
&=& \frac{1}{4}\left [ 1 + A\alpha_\Lambda\cos\theta_i^* 
+ B\alpha_\Lambda\cos\theta_j^* +C\alpha_\Lambda^2\cos\theta_i^*\cos\theta_j^* \right ],
\label{Eq:theta}
\end{eqnarray}
where 
\begin{eqnarray}
A=\frac{f_{\uparrow\uparrow}-f_{\downarrow\downarrow}+f_{\uparrow\downarrow}-f_{\downarrow\uparrow}}{f_{\uparrow\uparrow} + f_{\downarrow\downarrow} + f_{\downarrow\uparrow} + f_{\uparrow\downarrow}}, \notag \\
B=\frac{f_{\uparrow\uparrow}-f_{\downarrow\downarrow}-f_{\uparrow\downarrow}+f_{\downarrow\uparrow}}{f_{\uparrow\uparrow} + f_{\downarrow\downarrow} + f_{\downarrow\uparrow} + f_{\uparrow\downarrow}}, \notag \\
C=\frac{f_{\uparrow\uparrow}+f_{\downarrow\downarrow}-f_{\uparrow\downarrow}-f_{\downarrow\uparrow}}{f_{\uparrow\uparrow} + f_{\downarrow\downarrow} + f_{\downarrow\uparrow} + f_{\uparrow\downarrow}}.
\end{eqnarray}
The parameter $C$ is the $c_{ij}$ defined in Eq.~\ref{Eq:cnn}. Indeed, the parameter $A$ could be reduced to $(f_{\uparrow}-f_{\downarrow})/(f_{\uparrow}+f_{\downarrow})$ which represents the polarization of the first $\Lambda$ by definition, where $f_{\uparrow}=f_{\uparrow\uparrow}+f_{\uparrow\downarrow}$ ($f_{\downarrow}=f_{\downarrow\uparrow}+f_{\downarrow\downarrow}$) is the fraction of it in the spin state $\left | \uparrow  \right \rangle$ ($\left | \downarrow  \right \rangle$). Similarly, the parameter $B$ represents the polarization of the second $\Lambda$. It is reasonable to expect that parameter $C$ reflects the spin correlation between the first and second $\Lambda$. 
Mathematically, the $c_{ij}$ can be obtained by the integral of $\cos\theta^*_i\cos\theta^*_j$, it is equivalent to calculate the average $\langle \cos\theta^*_i\cos\theta^*_j \rangle$ in the experiment. 

Based on Eq.~\ref{Eq:cnn_prime}, we establish the connection between experimental observable and the $\Lambda$ spin correlation as used in the context of theoretical work: 
\begin{equation}
c'_{\Lambda\Lambda}=\frac{9}{\alpha^2_\Lambda}\langle \cos\theta^*_i\cos\theta^*_j \rangle - P_\Lambda^2.
\end{equation}
In experiment, the spin quantization direction (orbital angular momentum direction) is estimated from the event plane, which has a finite resolution. To correct for event plane resolution, similar to Ref.~\cite{STAR:2007ccu}, we project Eq.~\ref{Eq:theta} onto transverse plane, resulting in :
\begin{equation}
\frac{dN}{d\Delta\phi_i^*d\Delta \phi_j^*} = \frac{1}{16\pi}(1+A\alpha_\Lambda\sin\Delta \phi_i^* + B\alpha_\Lambda\sin\Delta \phi_j^*+\frac{C\alpha_\Lambda^2\pi}{4}\sin\Delta \phi_i^*\sin\Delta \phi_j^*),
\end{equation}
and $c'_{\Lambda\Lambda}$ can be calculated by
\begin{equation}
c'_{\Lambda\Lambda}=\frac{64}{\pi^2\alpha_\Lambda^2}\langle\sin\Delta\phi_i^*\sin\Delta \phi_j^* \rangle - P_\Lambda^2,
\label{Eq:FreeStyle}
\end{equation}
where $\Delta \phi_i^*$ ($\Delta \phi_j^*$) is the azimuthual angle relative to the reaction plane for proton $i$ ($j$) in first (second) $\Lambda$ rest frame. Considering a finite event plane resolution, Eq.~\ref{Eq:FreeStyle} can be expressed as
\begin{equation}
  c'_{\Lambda\Lambda}=\frac{64}{\pi^2\alpha_\Lambda^2}\frac{\langle\sin(\phi_i^*-\Psi_{\rm EP})\sin(\phi_j^*-\Psi_{\rm EP}) \rangle}{\langle \cos^2(\Psi_{\rm EP} - \Psi_{\rm RP}) \rangle} - P_\Lambda^2,  
\end{equation}
where $\Psi_{\rm EP}$ and $\Psi_{\rm RP}$ denote the event plane and the reaction plane, respectively. Note that
\begin{equation}
   \langle \cos^2(\Psi_{\rm EP} - \Psi_{\rm RP}) \rangle =   \frac{1 + \langle \cos(2(\Psi_{\rm EP} - \Psi_{\rm RP})) \rangle }{2},
\end{equation}
and $\langle \cos(2(\Psi_{\rm EP} - \Psi_{\rm RP})) \rangle$ can be determined from sub-event method in experiment~\cite{Poskanzer:1998yz}.

Building on the structure of balance function~\cite{STAR:2003kbb,Bass:2000az}, we propose a spin balance function to detect the difference of spin correlation between ``same sign'' pairs ($\Lambda$-$\Lambda$ or $\bar{\Lambda}$-$\bar{\Lambda}$) and ``opposite sign'' pairs ($\Lambda$-$\bar{\Lambda}$),
\begin{equation}
B(c'_{OS},c'_{SS}) = \frac{1}{2}(\frac{c'_{\Lambda\bar{\Lambda}} - c'_{\Lambda\Lambda}}{c'_{\Lambda\bar{\Lambda}} + c'_{\Lambda\Lambda}} + \frac{c'_{\bar{\Lambda}\Lambda} - c'_{\bar{\Lambda}\bar{\Lambda}}}{c'_{\bar{\Lambda}\Lambda} + c'_{\bar{\Lambda}\bar{\Lambda}}} ).
\end{equation}
The advantage of the above equation is that the detector effect and flavor independent correlation will be canceled, only additional correlation between $s$ and $\bar{s}$ (in $\Lambda$-$\bar{\Lambda}$ pairs) remains.

To summarize, in this paper, we investigated the correlation function used in a theoretical context, identified its connection to experimental observables, and outlined the procedure to measure it in experiments.  Additionally, we proposed a method for measuring  the  correlation using a spin balance function, which minimize the impact of detector and other trivial effects. This work will be instrumental in studying hyperon spin correlations and facilitating the understanding of spin-spin correlations and their implications for fluctuations in the strong force field. \\

\section*{Acknowledgment}
This work was inspired by the theoretical study \cite{Lv:2024uev} and benefited from discussions with X.-N. Wang, Z.-T. Liang, Q. Wang, X.-L. Sheng., and Z. Tu.

\bibliography{ref}

\end{document}